\documentstyle[apalike,11pt]{article}

\def\titel{Against Measurement? -- On the Concept of Information}
\def\be{\begin{equation}}
\def\ee{\end{equation}}
\def\ba{\begin{array}}
\def\ea{\end{array}}
\def\ket#1{ | #1 \rangle }
\def\bra#1{ \langle #1 | }
\def\braket#1#2{ \langle #1 | #2 \rangle }
\def\Bket#1{ \Big| #1 \Big\rangle }

\begin{document}
\sloppy

\thispagestyle{empty}

\begin{center}

{\Large\bf \titel}\footnote{
           Talk at the X-th Max Born Symposium ''Quantum Future'',
           Wroc{\l}aw, September 24 - 27, 1997\\
           (http://xxx.lanl.gov/abs/quant-ph/9709059)
           }

\vspace{7mm}

{\large Holger Lyre}\footnote{
       Institut f\"ur Philosophie,
       Ruhr-Universit\"at Bochum,
       D-44780 Bochum,
       Germany,\\
       e-mail: holger.lyre@ruhr-uni-bochum.de,
       http://www.ruhr-uni-bochum.de/philosophy/staff/lyre.htm
       }

\vspace{7mm}

{\large September 1997}

\vspace{11mm}

\begin{abstract}
In his last article {\em Against `Measurement'} \ {\sc J. S. Bell}
sums up his well known critique of the problem of
explaining the measurement process within the framework of quantum theory.
In this article I will discuss the measurement process by
analysing the concept of measurement from the epistemological
point of view and I will argue against {\sc Bell}
that it belongs to the preconditions of experience
to necessarily end up with a ''reduction of the wavefunction''.
I will consider the ''chain of reduction'' in detail --
from pure states of $\cal S \otimes A$ (system $\cal S$
and measuring apparatus $\cal A$)
via different kinds of mixtures to pure states of $\cal A(S)$.
It turns out that decoherence is not sufficient to explain reduction,
but that this can be done in terms of the concept of information
within a transcendental approach.
\end{abstract}

\end{center}


\vspace{9mm}

{\normalsize
\begin{quote}
{\Large Contents}\\

1 \ Introduction                                 \hfill 2

2 \ The Measurement Problem                      \hfill 2

3 \ The Chain of Reduction                       \hfill 3

4 \ Against Measurement? -- Some Interpretations \hfill 5

5 \ Information as a Transcendental Concept      \hfill 7

6 \ The Measurement Problem -- Revisited         \hfill 8
\end{quote}
}

\newpage


\section{Introduction}

Quantum theory is the `hard core' of physics today.
Its experimental success is overwhelming. But although the mathematical
framework and its pragmatical application is non-controversial,
it suffers from an obstinate interpretation problem:
what does the formalism really mean philosophically?
One part of this open question is the {measurement problem}.
In this article I first like to discuss the measurement problem
in terms of measurement theory and in comparison with
some actual interpretation approaches.
Since {\sc John S. Bell} in his famous article
{\em Against `Measurement'} \cite{bell90}
gives a very sophisticated (and also humorous) analysis
of the problems of quantum theory according to measurement theory,
I like to deal with some of his clever arguments -- including
the usage of his abbreviation
\begin{center}
{\em FAPP = for all practical purposes}.
\end{center}

The final aim is to stress the epistemological point of view
in discussing the measurement problem. This will be done
using a transcendental argument in the spirit of {\sc Kant}.
It turns out that in opposition to {\sc Bell}
the concept of information plays a key role in quantum theory.


\section{The Measurement Problem}
\label{problem}

The measurement problem is caused by the {\em universality}
of quantum theory, which means that there exists no inherent reason
not to apply it to any physical system or object.
In the theory of measurement the world is divided into distinct parts:
a system $\cal S$, on which the measurement is performed,
a measuring apparatus $\cal A$, and the `rest of the world' $\cal R$.
Because of its universality all parts must in principle
be describable by quantum theory.

Let us briefly review the steps of the measurement process.
I will denote the quantum states of $\cal S$
by $\ket{\psi}$ and the states of $\cal A$ by $\ket{\chi}$.
The observable $\hat A$ satisfies the eigenvalue equation
\be
\hat A \ket{\psi_i} = a_i \ket{\psi_i} .
\ee
To perform a measurement the {\sc Hilbert} space of $\cal S$
must be enlarged to the {\sc Hilbert} space
of the compound system $\cal S \otimes A$
by forming the tensor product which is spanned by the states
\be\label{S+A}
\ket{ \Phi } = \ket{ \psi \otimes \chi } = \sum_i c_i \ket{ \Phi_i }
\ee
with
\be\label{Phi_i}
\ket{ \Phi_i } = \ket{ \psi_i \otimes \chi } .
\ee
The tensor product also contains interference terms,
which represent the typical quantum correlations between $\cal S$
and $\cal A$. The state (\ref{S+A}) is a pure state with
a projection operator
\be\label{Phi_rein}
\hat P_{\Phi} = \ket{ \Phi } \bra{ \Phi } .
\ee
After the measurement interaction $\hat H_{int}$
between $\cal S$ and $\cal A$ we obtain
\be\label{S+A'}
\ket{ \Phi' } = e^{ i \hat H_{int} t  } \ \ket{ \Phi }
\ee
with states
\be
\Bket{ \Phi_i' } = \Bket{ \psi_i' \otimes \chi'(\psi_i) }
\stackrel{i.m.}{=} \Bket{ \psi_i  \otimes \chi'(\psi_i) }
\ee
instead of (\ref{Phi_i}).
In case of an ideal measurement we can replace
$\ket{ \psi' } \stackrel{i.m.}{=} \ket{ \psi }$.
Note that now the states $\Bket{ \chi'( \psi_i) }$
of the measuring apparatus are not independent from those of the system
$\ket{ \psi }$.

After the measurement interaction the compound system (\ref{S+A'})
must be separated into the subsystems $\cal S$ and $\cal A$ in order
to read $\cal A$. This is done by a cut.
Firstly, $\hat P_{\Phi'}$ transforms into the mixed states
of $\cal S$ and $\cal A$ with the density operators
$\hat\rho_{\psi}$ and $\hat\rho_{\chi}$
\be
\hat P_{\Phi'} \longrightarrow
\hat\rho_{\psi} = \sum_{i,k} w_{ik} \ket{ \psi_i } \bra{ \psi_k } .
\ee
Due to decoherence the interference terms will become immensely
small (but do not vanish exactly).

In principle the mixed state $\hat\rho_{\psi}$ allows an
infinite number of possible decompositions into states
$\ket{ \psi_i }$ of $\cal S$
(resp. into states $\Bket{ \chi( \psi_i) }$ of $\cal A$).
By picking out one special decomposition the mixed state
transforms into a mixture of states\footnote{In German:
''Gemisch'' (mixed state) vs. ''Gemenge'' (mixture of states)
\cite[p. 43]{heisenberg30}, \cite{suessmann58}. }
\be\label{mixture}
\hat\rho_{\psi}
\longrightarrow \hat\gamma_{\psi} = \Big\{ ( w_i, \hat P_{\psi_i} ) \Big\}.
\ee
As a last step the system is in a new state $\hat P_{\psi_i}$
\be
\hat\gamma_{\psi} \longrightarrow
\hat P_{\psi_i} = \ket{ \psi_i } \bra{ \psi_i } .
\ee
Likewise the measuring apparatus will show one result
and is therefore in a definite
pointer state $\hat P_{ \chi( \psi_i) }$.
Finally, at the end of the whole measuring act, the systems
$\cal A$ and $\cal S$ are in new pure states.


\section{The Chain of Reduction}
\label{chain}

As a result of the preceding section we see that strictly speaking
the so called `reduction of the wavefunction' has to be considered
as a {\em chain of reduction} of at least three steps:
\[
\ba{ccccccl}
 & & \hat\rho_{\psi} & \longrightarrow & \hat\gamma_{\psi} & \longrightarrow & \ \hat P_{\psi_i} \\
\hat P_{\Phi'} & {\displaystyle \nearrow \atop \displaystyle \searrow} \\
 & & \hat\rho_{\chi} & \longrightarrow & \hat\gamma_{\chi} & \longrightarrow & \ \hat P_{\chi(\psi_i)} \\
 & \mbox{\footnotesize (step 1)} & & \mbox{\footnotesize (step 2a)} & & \mbox{\footnotesize (step 2b)}
\ea
\]

Let us consider this chain step by step.

\paragraph{Step 1.}
Due to the measurement interaction and due to the coupling between
systems $\cal S$, $\cal A$, and $\cal R$ this step leads
to the cancellation FAPP of the interference terms
between $\cal S$ and $\cal A$,
i.e. the typical quantum correlations disappear (FAPP, but not exactly).
Thus, step 1 can be looked upon as a unitary
temporal development according to some wave equation.

\paragraph{Step 2.}
Due to the objectification the mixed state must be replaced
by a mixture $\hat\gamma$ according to (\ref{mixture}).
It is first of all important to note that
the quantum theoretical framework gives no possibility
to describe this step as a pure quantum process
(i.e. as a time evolution by a unitary operator).
{\sc Bell} points out that, logically speaking,
a system described by $\hat\rho$ is in a state
\begin{center}
{\em $\Psi_1 \Psi_1^*$ \ and \ $\Psi_2 \Psi_2^*$ \ and \ ... ,}
\end{center}
whereas $\hat\gamma$ refers to a state
\begin{center}
{\em $\Psi_1 \Psi_1^*$ \ or  \ $\Psi_2 \Psi_2^*$ \ or  \ ... .}
\end{center}
Although step 2 is certainly not involved in step 1 it is indeed
astonishing that this problem is very often not discussed in the common
measurement theory -- or as {\sc Bell} puts it:
{\em ''The idea of elimination of coherence, in one way or another,
implies the replacement of `and' by `or', is a very common one among
the solvers of the `measurement problem'. It has always puzzled me.''}
\cite[p.~36]{bell90}.

Step 2 can be split into two logical steps 2a and 2b
({\sc Bell} makes no explicit distinction between them):

\paragraph{Step 2a.}
A mixed state $\hat\rho$ represents a class of equivalent mixtures
$\hat\gamma$ -- such that from the mathematical point of view
no particular mixture is distinguished.
Step 2a stresses the point that one particular has to be choosen.

\paragraph{Step 2b.}
At the end of the whole measuring act the apparatus must show
exactly one result (on a display for instance) -- what else could
be the meaning of the term `measurement'?
Step 2b stresses the point that any satisfying theory of measurement
must explain the final occurrence of exacly {\em one} measuring outcome
out of the set of the many possible ones.

To make a distinction between steps 2a and 2b could be confusing.
The mixture $\hat\gamma$ can be considered as a statistical description
of different wavefunctions $\Psi_i$ -- such as the density matrix
in classical statistical thermodynamics.
This means, ontologically speaking,
that the system already exists in an actual state $\Psi_i$,
but it is not known (only with probabilities) to the observer.
Thus, the so called {\em ignorance interpretation} is valid
for $\hat\gamma$ but not for $\hat\rho$.
From this viewpoint step 2b would just describe the reading of the
measuring device by an observer and would be of no philosophical interest.
But it should be emphasized that one cannot distinguish between
$\hat\rho$ and $\hat\gamma$ by any observation.
With respect to this the distinction between step 2a and 2b indicates
the logical difference between `picking out a certain mixture $\hat\gamma$'
and `actually being in one certain state'
(the final pointer state of $\cal A$ for instance).


\section{Against Measurement? -- Some Interpretations}
\label{against}

What do the most prominent interpretation programs of quantum theory
today answer to the above steps in the chain of reduction?
I will briefly discuss some of them.

\paragraph{The Decoherence Program.}
      Nowadays, the concept of decoherence to explain step 1
      is very popular with physicists. It seems indeed very useful
      to scrutinize the conditions and orders of magnitude under which
      quantum systems decohere, but one has to keep in mind that
      decoherence is essentially FAPP.
      The question arises wether a FAPP description will be a
      satisfactory explanation. {\sc Bell} was obviously not satisfied
      -- although he and nobody, I presume, would deny the
      {\em ''... absence FAPP of interference between macroscopically
      different states''} \cite[p.~36]{bell90}.
      All in all decoherence is sufficient to explain step 1,
      whereas step 2 is by no means explained.

\paragraph{The {\sc Bohm-de~Broglie} Program or the Hidden Variable Program.}
      It is the first of {\sc Bell}'s favourites in his article.
      Of course a hidden variable argument is an objection which can
      always be raised: there could be something we do not know today!
      But, as a consequence of {\sc Bell}'s own invention
      -- his famous inequalities \cite{bell65} --
      {\em local} hidden variables are nowadays experimentally excluded.
      But does a theory with non-local hidden variables really show
      a conceptual difference to common quantum theory?
      New arguments support the idea that this is indeed not the case
      \cite{englert_etal92} -- would {\sc Bell} have believed in them?

\paragraph{The {\sc Ghirardi-Rimini-Weber} Model.}
      This is the second of {\sc Bell}'s favourites in his article.
      Its starting point is the {\em ad-hoc-assumption}
      that the wavefunction will 'collapse' after a given small and
      stochastic time interval by a spontaneous localization process.
      The parameters of the model are choosen for it to be in good
      correspondence with the ordinary quantum predictions.
      The model `explains' step 1 in the bandwidth of
      its parameters -- i.e. very well, but FAPP.
      Insofar the spontanous processes are {\em stochastic}, the central
      questions behind step 2a and, most of all, 2b remain
      unanswered -- thus the situation resembles the decoherence program.
      The question persists how a mere ad-hoc-model could be a
      satisfactory explanation of the deep problem of quantum
      measurement.

\paragraph{The Many Worlds Interpretation.}
      This interpretation leads essentially to {\em quantum cosmology},
      i.e. the reduction problem is considered for the universe
      as a whole. It does `explain' -- in a very broad-minded
      meaning of this word -- step 2,
      but it does not explain step 1.
      Consider for instance the following open questions:
      At what time steps does the `branching' of the universes occur
      and how many universes do occur at each time step?
      Today many authors combine the idea of decoherence
      with the concept of many worlds.
      But nevertheless the `ontological costs' of assuming many
      universes are very high!
      Should this really be the right answer? In any case,
      even {\sc Bell}, also in combination with the {\sc Bohm-de~Broglie}
      theory, {\em ''... did not like it''} \cite{bell76}.

\paragraph{The Copenhagen Interpretation (CI).}
This is the orthodox interpretation of quantum theory,
as far as it refers to its founders {\sc Werner Heisenberg} and,
most of all, {\sc Niels Bohr}.
Since there does not exist a kind of `codification' of CI,
there is still a certain confusion about its basic concepts.
I like to propose the following as the central assumption of CI:
{\em The outcomes of measurements must be described in classical terms,
i.e. the measuring apparatus must be described classically.}\footnote{
Compare {\sc Bohr} who emphasized that
{\em ''... however far the phenomena transcend the scope
of classical physical explanation, the account of all evidence
must be expressed in classical terms''} \cite[p.~209]{bohr49}.}
What is the meaning of this assumption?
It certainly means not that apparatuses and measuring devices are
non-quantum systems. But it means that the apparatuses must
{\em necessarily} be described classically in order to give
an appropriate description of the outcomes of a measurement.
`Appropriate' here means that the outcomes have to be communicable
and understandable to each observer. E.g., the idea that the pointer
of a device should be in a superposition of different pointer positions
is obviously senseless and non-communicable.
In this sense experimental data must be described classically.

Moreover, CI can be read as just offering the {\em minimal semantics}
to quantum theory, i.e. semantics which is necessary in order to apply
the theory to reality \cite{goernitz+cfw91b}.
Generally speaking, a physical theory contains two parts:
the mathematical structure of the formalism and the related physical
concepts. The minimal semantics of CI is:

\begin{tabular}{l|l}
Mathematical structure       & Physical concepts \\
\hline
{\sc Hilbert} space $\cal H$    & object (system) \\
(topological) tensor product  ${\cal H}_1 \otimes {\cal H}_2$
                                & composition of the objects 1 and 2 \\
(self-adjoint) operator         & observable \\
unitary U(1) transformation     & temporal development\\
vector  $\ket{\Psi} \in \cal H$ & state of an object       \\
scalar  $p=\Big| \braket{\Psi_i}{\Psi_f} \Big|^2$
        & probability for the transition of state $i$ to $f$
\end{tabular}

\vspace{3mm}

The last row refers to the central concept of CI: {\em probability}.
The scalar product of two states gives the amplitude of the
transition probability between them. Thus, the {\sc Hilbert} space
is provided with a probability metric.

A further remark should be made:
probability can be seen as the mathematical quantification of `possibility'.
Interestingly, the many worlds interpretation is in some sense not richer
than CI, since there exists a one-to-one terminological mapping between
both interpretations: the many world interpreters have simply
replaced the term `possibility' by the term `world'.
It seems that `many worlds' is just a fancy way of saying
something very trivial, namely `many possibilities' (i.e. probability).

But a crucial question remains:
how can a probability theory be understood without the concept of measurement?
How can physics be {\em ''against measurement''}?


\section{Information and the Transcendental Approach}

As shown in section \ref{chain}, the `reduction of the
wavefunction' is logically more than one single step.
CI gives a necessary minimal semantics
to quantum theory, but the alternative interpretive attempts
in section \ref{against} do not go beyond it to provide
a satisfactory understanding of the entire chain of reduction.
Thus, maybe a philosophical invention would be helpful:
the transcendental point of view.
As I like to argue in the following, this will not contradict CI,
but it will extend the
{\em epistemological aspect} of quantum theory.

At the end of section \ref{against} probability was identified as
the central concept of quantum theory -- at least according to CI,
its minimal semantics.
''Against {\sc Bell}'' I like to stress the point that in any probability
theory one can neither renounce the concept of measurement,
nor the concepts microscopic, macroscopic, reversible,
irreversible, observable, or observer -- some of the
{\em ''... worst terms''} in {\sc Bell}'s understanding.
Moreover, I like to propose {\em information}
-- decidedly against {\sc Bell} a really {\em good word} --
as the central concept behind it all \cite{lyre98}.
In view of the main question, what the quantum
wavefunction $\Psi$ really stands for,
the following basic assumption should be made
\begin{center}
$\Psi$ \ {\em is} \ information.
\end{center}

What are the reasons for believing this?
Probability means -- due to a logarithm - the same as the
syntactic aspect of information $I_{syn} \sim - \ln p$.
There also exist a semantic and a pragmatic aspect.
Thus, probability can be seen as a sub-concept of information.
Information, moreover, involves the terms subject and object.
Let us try to answer {\sc Bell} (1990, p.~34):
{\em ''Information? Whose information?''}
Information for any subject with conceptual and empirical competence!
{\em ''Information about what?''}
Information about empirically knowable objects, which are constituted
just by the information which can be gained from them!
In short:
Objects are constituted by information which is available to subjects.

In order to make the above answers plausible the direction of arguments
must be changed. This can be done by using a transcendental argument
in the manner of {\sc Immanuel Kant}:
the foundations of empirical science
are based on the preconditions of experience \cite{kant_KrV} -- and,
certainly nowadays, the foundations of empirical science are
the foundations of quantum theory \cite{drieschner79}.
The key idea is that since experience, and moreover empirical science,
is obviously possible and successful,
certain preconditions of experience must hold,
which, in the last analysis, make experience possible.
These preconditions will of course never fail in experience -- per
definition they can never become empirically falsified.
The crucial question then is: What are good candidates for
preconditions of experience? I like to propose just these two:
{\em distinguishability} and {\em temporality}. Why?
Whithout the possibility of making distinctions we could never be
able to have speech, concepts, and communicable thoughts.
If science is possible distinguishability is one of its most
rudimentary methodological preconditions.
Further on, without implicitly using the already known difference
between past and future we could never give any meaning to the word
experience -- or as {\sc Carl Friedrich von Weizs\"acker} puts it:
{\em ''... experience means to learn from the past for the future, then
any empirical science presupposes an understanding of past and future''}
\cite{goernitz+cfw91b}. This understanding may be called temporality.

Translated into an information theoretic language the difference between
past and future can be expressed as the difference between
{\em actual} and {\em potential information}.
More detailed than the above statement we can say
that the wavefunction or, in general, density operators
represent potential information.
Still more detailed, it is {\em quantum information},
since quantum bits are composed by the tensor product of two dimensional
{\sc Hilbert} spaces, i.e. quantum bits are indistinguishable
in contrast to bits of classical potential information.
In an earlier paper \cite{lyre97} I have argued that the so called
{\em complete concept of information}, i.e. the syntactic,
semantic, pragmatic, and temporal aspect of information,
can conceptually be deduced from distinguishability and temporality alone.
In this sense information can be based transcendentally,
i.e. concerning the preconditions of experience.


\section{The Measurement Problem -- Revisited}

What are the advantages in expanding CI strictly towards
an information theoretic interpretation of quantum theory?
In this last section I like to revisit the measurement problem
-- and especially the chain of reduction.
I will do this in a list of eight theses:

\begin{enumerate}

\item {\em Thesis. The chain of reduction is not a chain of physical
      interaction steps but of methodological ones.}
      The usual aim of measurement theory is to describe
      the measurement process as a quantum physical process
      by itself, i.e. to describe it as an interaction
      between $\cal S$ and $\cal A$.
      Surely, first of all the physical interaction $\hat H_{int}$
      -- as expressed in (\ref{S+A'}) -- establishes the measuring act.
      But this has nothing to do with the problem of reduction!
      Otherwise there should exist a unitary transformation
      for each step of the chain of reduction.
      But as analysed in section \ref{chain}, step 1 can only
      be considered as a time development FAPP,
      whereas step 2 certainly cannot.
      For the whole procedure described in section \ref{problem}
      we should better speak of measuring `act'
      instead of `process' (to speak with {\sc Bell}:
      `process' is a bad word at this stage).

\item {\em Thesis.} According to step 1: {\em
      Quantum theory does not predetermine the cut between
      $\cal S$ and $\cal A$, nevertheless the cut is necessary
      in order to apply quantum theory to reality}.
      Because of the universality of quantum theory,
      any system can in principle be described by quantum theory
      and, consequently, each measuring apparatus as well.
      As it was explained in section \ref{against}
      this is thoroughly compatible with CI.
      But in CI it is clearly seen
      that the universality of quantum theory is in a certain
      contradiction to its meaning as a theory of empirical science.
      Therefore from the CI viewpoint the idea of a wavefunction of the
      universe is a physically senseless extrapolation of the
      mathematical formalism.\footnote{Compare {\sc Heisenberg}:
      {\em ''... wenn man das ganze Universum in das System
      einbez\"oge -- dann ist ... die Physik
      verschwunden und nur noch ein mathematisches Schema geblieben''}
       \cite[p. 44]{heisenberg30},
      {\em ''... if the whole universe were to be included into the system
      then physics would vanish and just a mathematical scheme remains''}
      (translation by the author).}
      Is is, in principle, not forbidden to apply quantum
      theory to the world as a whole,
      but this would be no information for anybody
      since no subject is left.
      Thus, for any measurement the cut is a necessary precondition.

\item {\em Thesis.} According to step 2a: {\em
       The measuring apparatus must be suitable as such.}
       E.g., the pointer states must be orthogonal. Once a system is
       chosen to act as a measuring apparatus the decomposition of the
       density operator into the spectrum of the pointer variable
       is fixed. This is involved in the CI's central assumption
       of describing the measuring apparatus classically.

\item {\em Thesis.} According to step 2b: {\em
       Measurements must lead to irreversible facts.}
       Irreversibility in this context characterizes any documents
       of the past, e.g. pointer devices, printer outputs,
       computer memories, human brain states... .
       Facticity means that a measurement can lead to one and only
       one outcome. Thus, facts are classical (at least FAPP).

\item {\em Thesis.
      `Measurement' is a necessary term in any empirical science.
      It is related to information as the key concept of quantum theory.}
      Experience means to learn from the facts of the past
      for the possibilities of the future.
      In empirical science this will be done by measurements.
      Thus, in empirical science the term `measurement'
      is methodologically irreducible.
      Quantum physics, the hard core of empirical science,
      describes possibilities of the future in terms of potential
      information (''$\Psi$ is information'').
      Potential information is information which can be gained,
      i.e. can become actual, if a measurement is performed.

\item {\em Thesis. A measurement represents the transition from
      potential to actual information.}
      This thesis can mathematically be quantified.
      Quantum information is measured in terms of
      the entropy of the density operator
      $S \left[\hat\rho \right] = - k_B \ Sp \, ( \hat\rho \ \ln \hat\rho )$.
      A pure state contains no potential information,
      i.e. the initial as well as the final state of the chain of
      reduction represents $S\left[\hat P\right]=0$ bit.
      During step 2 the potential information amounts
      $S\left[ \hat\gamma \right] > 0$ bits.
      Since a measuring act represents the transition from
      potential to actual information,
      the reduction is a presupposition of the measurement.

\item {\em Thesis. The subject is an irreducible element in any
      empirical science.}
      According to the logic of the transcendental argument
      subjects in empirical science must be equipped at least
      with conceptual and empirical competence -- a
      {\em ''PhD''} \cite[p.~34]{bell90}
      is of course not a necessary precondition of experience!
      It should be noted that being a subject in this sense
      and being conscious is not necessarily the same.
      Thus, the assertion is not the reduction taking place
      in the consciousness of the observer
      as in the {\sc London-Bauer} or {\sc Wigner} approach,
      since in these approaches human consciousness is excluded
      from (quantum) physical description.
      But this contradicts the universality of quantum theory.
      The idea of the proposed transcendental approach is not
      to exclude anything from quantum theory, i.e. any subject can in
      principle be described by quantum theory -- but then, of course,
      as an object for and from another subject.
      This is the key point of the argument: subjects can be described
      as objects, but not all of them at the same time!
      Physics without any subjects would be meaningless.
      In that sense the subject is irreducible.

\item {\em Thesis. Physics is essentially FAPP, but `FAPPness'
       is no sufficient explanation to the measurement problem.}
      Objects are constituted by information which exists for subjects.
      Any (object) information presupposes a certain semantics under
      which the information can be understood. But for the same reason
      the information invested in the semantics needs other semantics
      before and so on. In a finite world, constituted by a huge
      but finite amount of information this leads to an inherent
      circularity \cite{lyre97}.
      Therefore, empirical science is by no means as exact as its
      mathematical framework suggests.
      Strictly speaking there exist no isolated quantum objects.
      Quantum theory is a holistic theory which, in the empirical
      application, must necessarily be FAPP.\footnote{Recent
      research even shows the universality of quantum theory
      leading to problems of self-referentiality for inner observers.
      E.g., an inner observer $\cal A$ cannot distinguish between
      a pure state and a mixture of $\cal S \otimes A$,
      i.e. they are indistinguishable FAPP. Moreover,
      self-referentiality seems to be connected to incompleteness
      of quantum physics \cite{breuer97}, \cite{mittelstaedt98}.}
      But `FAPPness' is no sufficient explanation to the measurement
      problem. It explains step 1, whereas my proposal is that
      step 2 should be seen under the transcendental approach.

\end{enumerate}

All in all the measurement `problem' could be soluble or even vanishes,
if it is not seen as a problem on the intrinsic-physical level,
i.e. described as a physical interaction process,
but on the meta-level, i.e. seen on the basis of the
methodological and epistemological presuppositions of physics.
Thus, {\sc Bell}'s list of bad words in fact appears to be
a list of {\em necessary} terms of any empirical science
-- most of all the terms `measurement' and `information'.

\vspace*{12mm}


\newpage

\bibliographystyle{apalike}

\end{document}